\documentclass[12pt]{article}
\usepackage{amsfonts}
\usepackage{amsmath}
\usepackage{latexsym}
\usepackage{amssymb}
\usepackage[mathcal]{euscript}
\usepackage{amscd}

\def\ben{\begin{equation}}
\def\een{\end{equation}}
\def\bea{\begin{eqnarray}}
\def\eea{\end{eqnarray}}
\input amssym.def
\input amssym.tex
\def\bea{\begin{eqnarray}}
\def\eea{\end{eqnarray}}
\input amssym.def
\input amssym.tex
\input{tcilatex}
\begin{document}

\title{A maximum magnetic moment to angular momentum conjecture}
\author{John D. Barrow$^{1}$ and G. W. Gibbons$^{1,2,3}$ \\
$^{1}$DAMTP, Centre for Mathematical Sciences,\\
University of Cambridge,\\
Wilberforce Rd., Cambridge CB3 0WA, UK\\
$^{2}$LE STUDIUM, Loire Valley Institute for Advanced Studies,\\
Tours and Orleans, France \\
$^{3}$Laboratoire de Math\'{e}matiques et de Physique \\
Th\'{e}orique, Universit\'{e} de Tours, France \\
}
\date{\today }
\maketitle

\begin{abstract}
Conjectures play a central role in theoretical physics, especially those
that assert an upper bound to some dimensionless ratio of physical
quantities. In this paper we introduce a new such conjecture bounding the
ratio of the magnetic moment to angular momentum in nature. We also discuss
the current status of some old bounds on dimensionless and dimensional
quantities in arbitrary spatial dimension. Our new conjecture is that the
dimensionless Schuster-Wilson-Blackett number, $c\mu /JG^{\frac{1}{2}}$,
where $\mu $ is the magnetic moment and $J$ is the angular momentum, is
bounded above by a number of order unity. We verify that such a bound holds
for charged rotating black holes in those theories for which exact solutions
are available, including the Einstein-Maxwell theory, Kaluza-Klein theory,
the Kerr-Sen black hole, and the so-called STU family of charged rotating
supergravity black holes. We also discuss the current status of the Maximum
Tension Conjecture, the Dyson Luminosity Bound, and Thorne's Hoop Conjecture.
\end{abstract}

\input amssym.def \input amssym.tex

\section{Introduction}

Regardless of what one thinks of the debate concerning the relative merits
of the traditional Baconian or inductionist, versus Bayesian or Popperian,
viewpoints about the nature of science, few would disagree that making
precisely stated conjectures or exhibiting counter-examples has an important
place in theoretical physics. In making such conjectures it is important to
bare in mind that although it is frequently convenient to adopt units well
suited to practical aspects of the subject being discussed, any physically
meaningful statement must be independent of an arbitrary choice of units. In
fact, adopting an appropriate set of \textquotedblleft natural
units\textquotedblright\ can afford insights which may be otherwise
obscured. In this paper we are led in section 2, by our consideration of
natural units for physical quantities which are independent of Planck's
constant, to conjecture new fundamental bounds on dimensionless quantities
in classical gravitation, in particular that there is an upper bound on the
magnetic moment to angular momentum ratio. In section 3, we verify that such
a bound holds for charged rotating black holes in those theories for which
exact solutions are available, including the Einstein-Maxwell theory,
Kaluza-Klein theory, the Kerr-Sen black hole, and the so-called STU family
of charged rotating supergravity black holes. We discuss the current status
of the Maximum Tension Conjecture in section 4, the Dyson luminosity bound
in section 5, and new approaches to Thorne's Hoop Conjecture in section 6.

\section{Units and dimensional analyses}

Natural units were first introduced into physics and metrology by George
Johnstone Stoney at the British Association Meeting in 1874, in an attempt
to cut through the proliferation of parochial units of measurement spawned
by the industrial revolution and the expansion of Victorian engineering and
commerce \cite{GJS, con}. He sought to devise units that, unlike feet and
horsepower, avoided any anthropomorphic benchmark, and made no use of
changing parochial standards, like days or standard weights. A similar
universal approach had also been advocated by Maxwell in 1870, who suggested
that constants be founded on atomic or optical standards \cite{max, con}. He
also saw a new opportunity to promote his prediction of a new elementary
particle, which he first dubbed the 'electrolion' in 1881 and then renamed
the 'electron' in 1894, carrying a basic unit of electric charge, $e$, whose
numerical value he predicted using Faraday's Law and Avogadro's Number. The
electron was subsequently discovered by Thomson in 1897, and Stoney remains
the only person to have successfully predicted the numerical value of a new
fundamental constant of physics.

\subsection{Stoney Units}

In response to a challenge from the British Association to reduce or
organise the plethora of special units that had sprung up to service the
industrial revolution and Britain's trading empire, in 1874 Johnstone Stoney
first introduced a system of "natural units" of mass, length and time using
the speed of light, $c$, the Newtonian gravitational constant, $G,$and his
proposed electron charge, $e$, \cite{JDB1, BT, OH}. Stoney's natural units
were

\begin{equation}
M_{S}=\left( \frac{e^{2}}{G}\right) ^{1/2},\text{ }L_{S}=\left( \frac{Ge^{2}%
}{c^{4}}\right) ^{1/2},\text{ }T_{S}=\left( \frac{Ge^{2}}{c^{6}}\right)
^{1/2}.  \label{st}
\end{equation}%
These were the first natural units. However, we should note that in those
days before the theory of special relativity, the speed of light, $c$, did
not possess the absolute status that it would later assume and $e$ was still
just a hobby-horse of Stoney's (for some context see the history ref. \cite%
{whit}).

\subsection{Planck Units}

In 1899, a similar idea was introduced by Max Planck \cite{MP} to create
another set of natural units based on $c,G,$ and $h,$ the quantum constant
of action that bears his name. They differ from Stoney's units by a factor $%
\frac{1}{2\pi }(\frac{e^{2}}{\hbar c})^{1/2}$ -- the square root of the fine
structure constant divided by $2\pi $. These units are now commonplace in
physics and cosmology and they define units of mass, length and time that
combine relativistic, gravitational and quantum aspects of physics:

\begin{equation}
M_{Pl}=\left( \frac{hc}{G}\right) ^{1/2},\text{ }L_{Pl}=\left( \frac{Gh}{%
c^{3}}\right) ^{1/2},\text{ }T_{Pl}=\left( \frac{Gh}{c^{5}}\right) ^{1/2}.
\label{Pl}
\end{equation}

However, innumerable related Planck units may be constructed for other
physical quantities in any number of space dimensions by dimensional
analysis. Those involving thermal physics can be included by adding the
Boltzmann constant, $k_{B}$, to $G,c$ and $h$. Some of the Planck units are
especially interesting for classical physics if they do not contain Planck's
constant. This signals that they are purely classical in origin and may
highlight a limiting physical principle. This is trivially so for the Planck
unit of velocity, $V_{Pl}=c$, but less obvious for the Planck units of force 
$F_{Pl}=c^{4}/G$ and power $P_{Pl}=c^{5}/G$ which are strongly suspected to
be maximal quantities in classical physics. It has been conjectured \cite%
{Gibbons:2002iv,Schiller1,Schiller2,Schiller3,JB2} \footnote{%
for an earlier anticipation of this idea but based on a different physical
motivation see \cite{Sabbata}}that in general relativity (with and without a
cosmological constant) there should be a maximum value to any physically
attainable force (or tension) given by 
\begin{equation}
F_{\max }=\frac{c^{4}}{4G}\,,  \label{1}
\end{equation}%
where $c$ is the velocity of light and $G$ is the Newtonian gravitational
constant. For possible relations to the holographic principle and to quantum
clocks, see \cite{Bolotin:2015uwa,Bolotin:2016roy,wig}.

\subsection{De Sitter units}

If one believes that the observed acceleration of the scale factor of the
universe \cite{Riess:1998cb} is due to a classical cosmological constant $%
\Lambda $ rather than some form of slowly-evolving 'dark energy', with
time-dependent density, then a set of absolute de Sitter units of mass,
length, and time can be introduced: 
\begin{equation}
M_{dS}=c^{2}G^{-1}\Lambda ^{-\frac{1}{2}}\,,L_{ds}=\Lambda ^{-\frac{1}{2}%
}\,,T_{ds}=c^{-1}\Lambda ^{-\frac{1}{2}}\,.  \label{ds}
\end{equation}%
In these units $c^{4}/G$ is still the unit of force and the upper bound (\ref%
{1}) still appears to hold \cite{JB2}.

\subsection{Fundamental principles and dimensions}

We referred above to 'limiting principles',\ or what are sometimes called
'impotence principles'. In \cite{Gibbons:2002iv} the phrase 'maximum tension
principle'\ was used in the usual sense of 'fundamental principles', that is
general statements expected to be true of all viable theories and which may
follow as a valid consequences of a precisely formulated mathematical
statement within any well-defined mathematically theory. Such principles may
have heuristic value in motivating and formulating a theory, but cannot be
used in themselves to define a theory. For example 'Heisenberg's Uncertainty
Principle'\ is an elementary theorem in wave mechanics but is insufficient
in itself to define wave mechanics. Moreover, it not only rests heavily on
translation invariance, but may not hold in more general quantum mechanical
theories, such as relativistic quantum field theory, in which the notion of
a position observable is problematic. Other examples in general relativity
include 'Mach's principle' \footnote{%
For an incisive account of many inequivalent formulations this can be given
see \cite{Bondi:1996md}.} 'equivalence principles. and Thorne's 'hoop
conjecture' (to which we return below). Other 'principles', like the
'cosmological principle' may be simplifying symmetry assumptions, or
approximations, that cannot be precisely true in reality, or straightforward
methodological principles, like the 'weak anthropic principle', or various
variational principles.

The maximum force conjecture gives rise to the closely related conjecture 
\cite{Massa} that there is a maximum power defined by 
\begin{equation}
P_{\max }=cF_{\max }=\frac{c^{5}}{4G},  \label{2}
\end{equation}%
the so-called Dyson luminosity \cite{dys}, or some multiple of it (to
account for geometrical factors that are $O(1)$). This will be treated in
detail in section 4.

We note that some of the non-quantum Planck units, like the velocity, $%
V_{Pl}=c$, are independent of the dimension of space but others, like $%
F_{Pl} $, are not, because in $N$-dimensional space the dimensions of $G$
are $M^{-1}L^{N}T^{-2}$. Thus, in $N$ dimensions the non-quantum Planck unit
is mass $\times $ (acceleration)$^{N-2}$, which is only a force when $N=3,$
as shown in ref. \cite{JB2}.

In this paper, we display another physically interesting non-quantum Planck
unit formed by the ratio of the magnetic moment of a body, $\mu $, to its
total angular momentum, $J$, and conjecture that classically all bodies
satisfy an inequality

\begin{equation}
\frac{\mu }{J}<\beta \frac{G^{1/2}}{c},  \label{3}
\end{equation}%
where $\beta $ is a numerical factor $O(1),$ and we explore the evidence for
this maximum bound. Unlike the Planck units of force and power, the Planck
unit for the ratio $\mu /J$ is independent of spatial dimension.

To show this, if we use unrationalised units the dimensions $[.]$ of
magnetic $Q$ and electric charge $\bar{Q}$ are the same and are given by the
inverse-square laws discovered by Michell and Priestley, \cite{Michell,
Priestley}, respectively, with 
\begin{equation}
\lbrack Q]=[\bar{Q}]=M^{\frac{1}{2}}L^{\frac{3}{2}}T^{-1}\,.  \label{4}
\end{equation}%
The dimensions of a magnetic moment $\mu $ are therefore 
\begin{equation}
\bigl [\mu \bigr ]=M^{\frac{1}{2}}L^{\frac{5}{2}}T^{-1}\,.  \label{5}
\end{equation}

Thus, the ratio of magnetic moment to angular momentum $J$ has dimensions 
\begin{equation}
\bigl [\frac{\mu }{J}\bigr ]=\bigl [\frac{G^{\frac{1}{2}}}{c}\bigr ]=\bigr [%
\frac{Q}{Mc}\bigr ]\,.  \label{6}
\end{equation}%
which is independent of Planck's constant $\hbar $. This property continues
to hold in $N$-dimensional space because there we have $%
[Q]=M^{1/2}L^{N/2}T^{-1}$, $[\mu ]=M^{1/2}L^{1+N/2}T^{-1}$ and $%
[J]=ML^{2}T^{-1}$.

The ratio 
\begin{equation}
Z\equiv \frac{Q^{2}}{GM^{2}}  \label{7}
\end{equation}%
may be regarded as the separation-independent ratio of the electrostatic
repulsion to the gravitational attraction between two identical bodies of
mass $M$ and charge $Q$. It has been claimed \cite{Kragh} that Z\"{o}llner
was the first person to recognise its significance and so one might call it
the Z\"{o}llner number. A famous, but now discredited, theory of Dirac's
predicting the time variation of the gravitation 'constant' $G\propto 1/t$,
with the age of the universe $t$, \cite{Dirac} was partly motivated by the
very small value of $Z$ when the mass $M=m_{e}$ and charge $Q=e$ of the
electron (or even the proton mass $m_{pr}$) are substituted. Thus, giving 
\begin{equation}
N=\frac{e^{2}}{Gm_{e}^{2}}\approx 3\times 10^{42},  \label{8}
\end{equation}%
which suggested to Dirac its possible equality (in some yet to be found
theory) with the square root of the total number of protons or electrons in
the visible universe, $c^{3}t/Gm_{e}\sim 10^{83}$, up to a factor $O(1)$. In
fact, the value of $N$ and its numerical proximity to the ratio of the
classical electron radius to the Hubble radius was first noticed by Weyl in
1919 \cite{weyl, BT} and the numerical 'coincidence' is anthropic because it
is equivalent to the statement that the present age of the universe is of
order the main sequence lifetime of a star \cite{dicke, BT}.

Classically, we have the Larmor relation 
\begin{equation}
\frac{\mu }{J}=\frac{Q}{2Mc}  \label{9}
\end{equation}%
where $M$ is the mass of a system with charge $Q$. More generally, we have 
\begin{equation}
\frac{\mu }{J}=g\frac{Q}{2Mc}  \label{10}
\end{equation}%
where $g$ is the gyromagnetic ratio. Famously, Dirac showed that for
electrons $g=2$, at least at lowest order in the fine structure constant $%
e^{2}/\hbar c,$ \cite{Dirac:1928hu}, and this value has some significance in
supersymmetric theories \cite{Ferrara:1992yc}.

After earlier suggestions made by Schuster \cite{Schuster} and Wilson \cite%
{Wilson}, Blackett \cite{Blackett} conjectured that all rotating bodies
should acquire a magnetic moment given by 
\begin{equation}
\frac{\mu }{J}=\beta \frac{G^{\frac{1}{2}}}{c}\,,  \label{11}
\end{equation}%
where the dimensionless Schuster-Wilson-Blackett number has $\beta \approx
O(1),$ and was once regarded as a possible universal constant. Although $%
\beta $ is found to be of order unity for a variety of rotating astronomical
bodies ranging from the earth, the sun, and a variety of stars, as a general
statement for macroscopic bodies, the Schuster-Wilson-Blackett conjecture
has fallen foul of astronomical data. Yet it remains of interest to enquire
whether it provides a natural upper bound for bodies with significant
gravitational self-energy.

Since, for electrons 
\begin{equation}
\beta =N^{\frac{1}{2}}\,,  \label{12}
\end{equation}%
no interesting bound holds for the elementary particles. However, it is of
interest to ask what is known about $\beta $ in Einstein-Maxwell and
supergravity theories, since for black holes there is typically an upper to $%
|Q|/G^{\frac{1}{2}}M$ of order unity. For Planck mass particles with charges
of order $e$, we find $\beta $ is not far from unity. Such objects can arise
in string theory, whose low-energy limit is supergravity theory, so this
further motivates the investigation that follows. \ \ \ \ \ \ \ \ 

\section{The Schuster-Wilson-Blackett Number for electrically charged
rotating black holes}

Brandon Carter first discovered \cite{Carter} that Kerr-Newman black holes
in Einstein-Maxwell theory have a gyromagnetic ratio equal to $2$: 
\begin{equation}
\frac{\mu }{J}=\frac{Q}{Mc}\,.  \label{13}
\end{equation}

Now, to avoid naked singularities, we require (if we assume the black hole
has no magnetic charge) 
\begin{equation}
GM^{2}\geq Q^{2}+\frac{J^{2}}{M^{2}}.  \label{14}
\end{equation}%
Thus, 
\begin{equation}
1\leq \frac{c^{2}\mu ^{2}}{GJ^{2}}+\frac{J^{2}}{GM^{4}},  \label{15}
\end{equation}%
and so we have the required bound: 
\begin{equation}
\big |\frac{\mu }{J}\big |\leq \frac{G^{\frac{1}{2}}}{c}\,.  \label{16}
\end{equation}%
Hence, we have $\beta <1$ for Kerr-Newman black holes. The literature on
extensions of Carter's result is quite large. A notable example \cite%
{Garfinkle:1990ib} is a detailed analysis of a current loop surrounding a
static black hole. As the loop moves towards the horizon the gyromagnetic
ratio smoothly interpolates between the classical value $g=1$ and the
Carter-Dirac value $g=2$.

It was shown by Reina and Treves \cite{Reina:1975rt} that any
asymptotically-flat solution of the Einstein-Maxwell equations obtained by
performing a Harrison transformation on a neutral solution must also have $%
g=2$. Furthermore, it has been shown \cite{Gibbons:1982fy,Gibbons:1982jg}
that, provided any sources obey the constraint that $G$ times the energy
density bounds the charge density, then all asymptotically-flat solutions of
the Einstein Maxwell equations, possibly with sources of the kind specified
which are regular outside a regular event horizon, obey the following
Bogomolnyi bound on the Z\"{o}llner number: 
\begin{equation}
Z=\frac{Q^{2}}{GM^{2}}\leq 1.  \label{17}
\end{equation}%
Combining this with Reina and Treves' result, implies 
\begin{equation}
\beta <1.  \label{18}
\end{equation}

\subsection{Kerr-Newman AdS black holes}

Using the notation of \cite{Aliev:2006tt}, and temporarily setting $G=c=1$,
we must distinguish the parameters $M,Q,J,$ in the spacetime metric from the
physical quantities. The latter are denoted by primes. From \cite%
{Gibbons:2004ai}, we introduce 
\begin{equation}
M^{\prime }\equiv \frac{M}{\Xi ^{2}}\,,\qquad J^\prime \equiv \frac{aM}{\Xi
^{2}}  \label{19}
\end{equation}%
where $\Xi \equiv 1-\frac{a^{2}}{l^{2}}$.

Aliev gives the physical charge as 
\begin{equation}
Q^{\prime }=\frac{Q}{\Xi }\,,  \label{20}
\end{equation}
He finds 
\begin{equation}
\mu ^{\prime }=\frac{Qa}{\Xi }\,,  \label{21}
\end{equation}%
so we have%
\begin{equation}
\frac{|\mu^{\prime }|}{|J^{\prime }|}=\frac{|Q|}{M}(1-\frac{a^{2}}{l^{2}})\,.
\label{22}
\end{equation}

Now, for a horizon to exist, we require 
\begin{equation}
\Delta _{r}=(1+\frac{a^{2}}{l^{2}})\Bigl(r^{2}-\frac{2Mr}{1+\frac{a^{2}}{%
l^{2}}}+\frac{Q^{2}+a^{2}}{1+\frac{a^{2}}{l^{2}}}\Bigr )+\frac{r^{4}}{a^{2}}
\label{23}
\end{equation}

to have at least one real root. A \emph{necessary} condition for this is
that the quadratic in the first term be negative. This requires 
\begin{equation}
\frac{|Q|}{M}<\frac{1}{\sqrt{1+\frac{a^{2}}{l^{2}}}}.  \label{24}
\end{equation}

Thus, we also require 
\begin{equation}
\frac{|\mu ^{\prime }|}{|J^{\prime }|}<\frac{1-\frac{a^{2}}{l^{2}}}{\sqrt{1+%
\frac{a^{2}}{l^{2}}}}.  \label{25}
\end{equation}

Now, 
\begin{equation}
(1-x)(1+x)=1-x^{2}\leq 1\,,\quad \Rightarrow \quad \frac{1-x}{\sqrt{1+x}}%
\leq \frac{1}{(1+x)^{\frac{3}{2}}}\leq 1\,,  \label{26}
\end{equation}

so 
\begin{equation}
\frac{|\mu ^{\prime }|}{|J^{\prime }|}<1\,.  \label{27}
\end{equation}%
Therefore, we have shown that $\beta <1$ for Kerr-Newman-AdS black holes. 
\footnote{%
Note that since Harrison transformations are not available when the
cosmological constant is non-vanishing, there is no analogue of the
Reina-Treves result with which to combine the Bogomolnyi bound of. \cite%
{Gibbons:1983aq} in this case.}

\subsection{Einstein-Maxwell-Dilaton black holes}

These have only been discussed for general dilaton-photon coupling constant $%
\alpha $ for the case of slow rotation \cite{Horne:1992zy,Shiromizu:1999bm}.
One has a uniqueness theorem for general $\alpha $, angular momentum,
electric and magnetic charges \cite{Yazadjiev:2010bj} provided that $\alpha
^{2}\leq 3$.

In the general slow-rotation case one finds that \cite{Horne:1992zy} 
\begin{equation}
J=\frac{a}{2}\bigl(r_{+}+\frac{3-\alpha ^{2}}{3(1+\alpha ^{2})}r_{-}\bigr )%
\,,\qquad \mu =aQ\,.
\end{equation}

If $a$ is small, then the mass $M$ and charge $Q$ are given by 
\begin{equation}
M=\frac{1}{2}\Bigl(r_{+}+\frac{1-\alpha ^{2}}{1+\alpha ^{2}}r_{-}\Bigr )%
\,,\qquad |Q|=\sqrt{\frac{r_{+}r_{-}}{1+\alpha ^{2}}}\,.
\end{equation}%
Since $r_{+}\geq r_{-}\geq 0,$ we have 
\begin{equation}
\frac{|Q|}{M}\leq \sqrt{1+\alpha ^{2}},
\end{equation}%
so that in accordance with the Bogolmolnyi bound of \cite{Gibbons:1993xt},
this gives 
\begin{equation}
M\geq \frac{|Q|}{\sqrt{1+\alpha ^{2}}}\,.  \label{Bog}
\end{equation}

We have 
\begin{equation}
\frac{|J|}{|\mu |}=\frac{1}{2}\sqrt{1+\alpha ^{2}}\Bigl(\sqrt{\frac{r_{+}}{%
r_{-}}}+\frac{3-\alpha ^{2}}{3(1+\alpha ^{2}}\sqrt{\frac{r_{-}}{r_{+}}}\Bigr)%
\,,
\end{equation}%
so, \textit{provided $\alpha ^{2}\leq 3$}, this gives 
\begin{equation}
\frac{|\mu |}{|J|}\leq \frac{1}{2}\sqrt{1+\alpha ^{2}}\leq 1\,.
\end{equation}

As pointed out in \cite{Horne:1992zy}, we can then obtain a gyro-magnetic
ratio: 
\begin{equation}
g=2-\frac{4\alpha ^{2}r_{-}}{(3-\alpha ^{2})r_{-}+3(1+\alpha ^{2})r_{+}}\,.
\end{equation}

\subsection{Kerr-Kaluza-Klein black holes}

The observational and theoretical failures of the old Schuster-Blackett
conjecture (\ref{11}) led some to resort to Kaluza-Klein theory (see \cite%
{Barut:1985uj}). Rotating charged black holes in this theory may be obtained
by boosting the neutral Kerr solution (sometimes referred to in this context
as a rotating 'black string') along the fifth dimension. If $v$ is the
velocity parameterizing the boost, and $a$ and $M_{s}$ the parameters of the
original Kerr solution, then in units in which $G=c=1$ \cite{Gibbons:1985ac},%
\cite{Frolov:1987rj}, we have 
\begin{eqnarray}
M=M_{s}\bigl(1+\frac{1}{2}\frac{v^{2}}{1-v^{2}}\bigr )\,,\qquad J &=&\frac{%
M_{s}a}{\sqrt{1-v^{2}}}  \notag \\
Q=M_{s}\frac{v}{1-v^{2}}\,,\qquad \mu &=&\frac{M_{s}av}{\sqrt{1-v^{2}}}
\label{KK1}
\end{eqnarray}

and the gyromagnetic ratio is $g=2-v^{2}$. Restoring units, we have 
\begin{equation}
\frac{|\mu |}{|J|}=\frac{G^{\frac{1}{2}}}{c}v,  \label{29}
\end{equation}%
and, remarkably, we see that $\beta =v/c\leq 1$.

We may also regard Kaluza-Klein black holes as Einstein-Maxwell-Dilaton
black holes with $\alpha =\sqrt{3}$. For the gyromagnetic ratios of
elementary particles in Kaluza-Klein theory and their comparison with black
holes, the reader may consult \cite%
{Hosoya:1983tc,Barut:1985uj,Gibbons:1985ac}.

\subsection{Kerr-Sen electrically-charged black holes}

These black holes satisfy the low-energy equations of motion of heterotic
string theory \cite{Sen:1992ua} and may be regarded as an
Einstein-Maxwell-Dilaton black hole with coupling constant $\alpha =1$.
According to \cite{Sen:1992ua}, the mass $M$, charge $Q$, angular momentum $%
J $ and magnetic dipole moment, $\mu ,$ are given by 
\begin{eqnarray}
M &=&\frac{m}{2}(1+\cosh \theta )\,,\qquad J=Ma  \notag \\
Q &=&\frac{m}{\sqrt{2}}\sinh \theta \,,\quad \mu =Qa\,.  \label{KK}
\end{eqnarray}%
where $m,a,\theta $ are parameters \footnote{%
Our $\theta $ is Sen's $\alpha $.}. Thus, we find 
\begin{equation}
\beta =\frac{|\mu |}{|J|}=\sqrt{2}\frac{\sinh \theta }{1+\cosh \theta }=%
\sqrt{2}\tanh \frac{\theta }{2}\leq \sqrt{2}\,.  \label{String}
\end{equation}%
and 
\begin{equation}
g=2\,.
\end{equation}

We also find a Bogomolnyi inequality (\ref{Bog}) with $\alpha ^{2}=1$ is
satisfied, that is, 
\begin{equation}
\frac{|Q|}{M}\leq \sqrt{2}.
\end{equation}

\subsection{ STU electrically charged black holes}

The electromagnetic properties of a more general family of 4-charged black
holes, which are solutions of the so-called $STU$ supergravity theory
(characterised by $S$, $T$, and $U$ dualities) are reviewed in \cite%
{Cvetic:2013roa}. These solutions depend upon 4 boost parameters, $\delta
_{i}$. If $c_{i}=\cosh \delta _{i}$, $s_{i}=\sinh \delta _{i}$, $\Pi
_{c}=c_{1}c_{2}c_{3}c_{4}$, $\Pi _{s}=s_{1}s_{2}s_{3}s_{4}$, $\Pi
_{c}^{1}=c_{2}c_{3}c_{4}\,\mathrm{etc}$, $\Pi _{s}^{1}=s_{2}s_{3}s_{4}\,%
\mathrm{etc}$, then according to \cite{Cvetic:2013roa} 
\begin{eqnarray}
\qquad M=\frac{m}{4}\sum_{i}\bigl(c_{i}^{2}+s_{i}^{2}\bigr )\,,\qquad J &=&ma%
\big(\Pi _{c}-\Pi _{s}\bigr) \\
Q_{i}=2ms_{i}c_{i}\,,\qquad \mu _{i} &=&2ma\bigl(s_{i}\Pi _{c}^{i}-c_{i}\Pi
_{s}^{i}\bigr )\,.  \label{CY}
\end{eqnarray}

Evidently 
\begin{equation}
4M \ge \sum_i |Q_i| \,.  \label{BOG}
\end{equation}

We also have 
\begin{equation}
\frac{1}{2}\frac{\mu _{i}}{J}=\frac{s_{i}\Pi _{c}^{i}-c_{i}\Pi _{s}^{i}}{\Pi
_{c}-\Pi _{s}}\,.
\end{equation}%
If we assume that $s_{i}>0$, $\forall \,i$, then 
\begin{equation}
\frac{1}{2}\frac{\mu _{i}}{J}\leq \frac{s_{i}(\Pi _{c}^{i}-\Pi _{s}^{i})}{%
c_{i}(\Pi _{c}^{i}-\Pi _{s}^{i})}\leq \tanh \delta _{i}\leq 1\,.
\label{BOUND}
\end{equation}

There are some special cases which coincide with solutions of the
Einstein-Maxwell-Dilaton theory.

\begin{itemize}
\item \textit{Einstein-Maxwell Black Holes}: $\delta _{1}=\delta _{2}=\delta
_{3}=\delta _{4}$, $Q_{i}=Q$, $\mu _{1}=\mu $.

Thus, from (\ref{CY}), we have 
\begin{eqnarray}
M=m\cosh {2\delta }\,,\qquad J &=&ma\cosh 2\delta , \\
Q_{i}=m\sinh 2\delta \,,\quad \mu _{i} &=&ma\sinh 2\delta ,
\end{eqnarray}%
and if we set $Q=Q_{i}$ and $\mu =\mu _{i}$ so that 
\begin{equation}
Q^{2}=\frac{1}{4}\bigl(Q_{1}^{2}+Q_{2}^{2}+Q_{3}^{2}+Q_{4}^{2}\bigr ),
\label{sum4}
\end{equation}%
then we find that $g=2$ and 
\begin{equation}
\frac{|\mu |}{|J|}=\tanh 2\delta =\frac{|Q|}{M}\leq 1\,.
\end{equation}

Evidently both the charge inequality (\ref{BOG}) and the dipole inequality ( %
\ref{BOUND}) are satisfied, the latter by a comfortable margin since for $%
x>0 $, 
\begin{equation}
\tanh 2x\leq 2\tanh x\,.
\end{equation}%
.

\item \textit{Kerr-Kaluza-Klein electrically charged black holes}: $\delta
_{2}=\delta _{3}=\delta _{4}=0$, $\delta _{1}=\delta $. 
\begin{eqnarray}
M=\frac{m}{4}(3+\cosh {2\delta }),\qquad J &=&ma\cosh \delta \\
Q_{1}=m\sinh 2\delta \,,\qquad \mu _{1} &=&2ma\sinh \delta
\end{eqnarray}%
Now, if $v=\tanh \delta $ then from (\ref{KK}) we have 
\begin{eqnarray}
M &=&\frac{M_{s}}{4}\bigl(3+\cosh 2\delta \bigr )\,,\qquad J=M_{s}a\cosh
\delta  \notag \\
Q &=&\frac{M_{s}}{2}\sinh 2\delta \,,\qquad \mu =M_{s}a\sinh \delta
\end{eqnarray}

Thus, $M_{s}=m$, $Q=\frac{1}{2}Q_{1}$ and $\mu =\frac{1}{2}\mu _{1},$ so
that we have 
\begin{equation}
Q^{2}=\frac{1}{4}Q_{1}^{2},  \label{sum1}
\end{equation}%
and we find that 
\begin{equation}
\beta =\frac{|\mu |}{|J|}=\tanh \delta \leq 1\,.
\end{equation}%
We also have 
\begin{equation}
M\geq \frac{1}{2}|Q|,
\end{equation}%
which is consistent with (\ref{Bog}) provided that $\alpha ^{2}=3$.

\item \textit{String Theory}: $\delta _{1}=\delta _{2}=\delta $, $\delta
_{3}=\delta _{4}=0.$

Now, we have 
\begin{eqnarray}
M=\frac{1}{2}m(1+\cosh 2\delta )\,,\qquad J &=&\frac{1}{2}ma(1+\cosh 2\delta
), \\
Q_{1}=Q_{2}=m\sinh 2\delta \,,\qquad \mu _{1}=\mu _{2} &=&2ma\sinh \delta
\cosh \delta \,,
\end{eqnarray}

and if we set $Q_{1}=Q_{2}=\sqrt{2}Q$ and $\mu _{1}=\mu _{2}=\sqrt{2}\mu ,$
so that 
\begin{equation}
Q^{2}=\frac{1}{4}(Q_{1}^{2}+Q_{2}^{2}),  \label{sum2}
\end{equation}%
we obtain 
\begin{equation}
\frac{|\mu |}{|J|}=\sqrt{2}\tanh \delta \,.
\end{equation}%
and 
\begin{equation}
\frac{|Q|}{M}=\sqrt{2}\tanh \delta \leq \sqrt{2}\,.
\end{equation}%
We also find that $g=2$ and obtain consistency with (\ref{Bog}) and
agreement with (\ref{String}) provided $\theta =2\delta $.
\end{itemize}

Note that for all these special cases, the conversion from the conventions
of \cite{Cvetic:2013roa} and standard (Gaussian) units is 
\begin{equation}
Q^2 = \frac{1}{4} \sum _i Q_i^2 \,.
\end{equation}

\section{The Dyson bound}

The importance of some multiple of $c^{5}/G$ in studies of gravitational
radiation appears to have first been noticed in a paper of Dyson \cite{dys}.
He observed, by a scaling argument, that the luminosity in gravitational
radiation of an orbiting binary star system according to Einstein's
linearised theory of gravitational radiation, must be a dimensionless
multiple of $c^{5}/G$, (see below for a more precise statement).
Subsequently, Thorne \cite{Thorne} introduced it into modern studies of
possible sources of gravitational radiation, linear or non-linear,
detectable on earth using current technology. Thorne's paper seems to have
introduced the idea of a Dyson bound\textit{\ }\cite%
{Sperhake:2011xk,Cardoso:2013krh}: a maximum possible luminosity in
gravitational waves \footnote{%
In reply to an enquiry by Christoph Schiller, Dyson replied on 14th Feb
2011: 'It is not true that I proposed the formula $c^{5}/G$ as a luminosity
limit for anything. I make no such claim. Perhaps this notion arose from a
paper that I wrote in 1962 with the title, \textquotedblleft Gravitational
Machines\textquotedblright , published as Chapter 12 in the book,
\textquotedblleft Interstellar Communication\textquotedblright\ edited by
Alastair Cameron, [New York, Benjamin, 1963]. Equation (11) in that paper is
the well-known formula $128V^{10}/5Gc^{5}$ for the power in gravitational
waves emitted by a binary star with two equal masses moving in a circular
orbit with velocity $V$. As $V$ approaches its upper limit $c$, this
gravitational power approaches the upper limit $128c^{5}/5G$. The remarkable
thing about this upper limit is that it is independent of the masses of the
stars. It may be of some relevance to the theory of gamma-ray bursts.'}.

Six years after \cite{dys}, Dyson wrote a short note, \cite{Dyson2}, posing
a question the answer to which was supplied by Hawking's famous area theorem 
\cite{Hawking:1971tu}. It seems reasonable therefore to suggest (see
footnote 9 of \cite{Abbott:2016bqf}) that $c^{5}/G,$ or some multiple of it,
be called \textquotedblleft one Dyson\textquotedblright . If one accepts
this, the maximum luminosity of GW150914 (or the orbital merger of any
equal-mass non-spinning black holes) is about $1$ milli-Dyson.

\section{The Maximum Tension Principle}

Independent of these considerations, in an article written for the
Festschrift celebrating the 60th birthday of the late Jacob Bekenstein \cite%
{Gibbons:2002iv}, it was conjectured that $c^{4}/4G$ was the maximum
possible tension or force in classical general relativity. Dimensionally,
this makes sense. The Einstein equations read 
\begin{equation}
R_{\mu \nu }-\frac{1}{2}g_{\mu \nu }g^{\alpha \beta }R_{\alpha \beta }=\frac{%
8\pi G}{c^{4}}T_{\mu \nu }\,.
\end{equation}%
The Ricci tensor $R_{\mu \nu }$ has dimensions $\mathrm{L}^{-2}$ and every
component of the energy-momentum tensor, $T_{\mu \nu }$, has dimensions
force per unit area and the Einstein constant $8\pi G/c^{4}$ has units of an
inverse force.

Some heuristic arguments in favour of this maximum tension conjecture were
given in \cite{Gibbons:2002iv} and the factor of $\frac{1}{4}$ justified by
reference to conical singularities and the requirement that the deficit
angles of cosmic strings do not exceed $2\pi $ radians. In fact, the deficit
angle is subject to a so-called Bogomolnyi bound \cite{Comtet:1987wi} in
this case. The extensions in the presence of a cosmological constant were
given in \cite{JB2} but as yet there exists no formal proof, or indeed
precise mathematical formulation. Further work on the maximum tension (or
force) principle may be found in \cite{Schiller1,Schiller2,Schiller3,Good}.
Earlier suggestions regarding a maximum force then came to light. In \cite%
{Sabbata}, the authors claimed that $c^{4}/4G$ is the maximal force allowed
in general relativity and in \cite{Massa} made the obvious maximal power
hypothesis that $c^{5}/4G$ is the maximum power allowed in nature. Neither
paper makes any reference to \cite{dys} or \cite{Thorne}. There are also
earlier (unseen) papers on this subject, \cite{Kostro:1999ue,Kostro:2000gw},
whose titles clearly indicates that the author had the same order of
magnitude for the maximal force and maximal power in mind \cite%
{Kostro:2010zz}.

\section{Thorne's Hoop Conjecture}

The proposed Dyson bound and the maximum tension principle resemble another,
as yet unresolved but possibly related, issue: how does one formulate in a
precise way Thorne's hoop conjecture? \cite{Thorne:1972ji} Recently, there
has been some progress in this direction.

In \cite{Gibbons:2009xm}, a precise candidate was proposed for the hoop
radius of an apparent horizon in terms of its Birkhoff invariant $\beta _{b}$%
. The conjecture was that every apparent horizon should satisfy 
\begin{equation}
\beta _{b}\leq 4\pi GM_{ADM}/c^{2},  \label{hoop1}
\end{equation}%
where $M_{ADM}$ is the ADM mass of the system. In \cite{Cvetic:2011vt},
considerable support was marshalled for (\ref{hoop1}) using known exact
solutions of supergravity theories. However, more recently, a counterexample
was constructed using time-symmetric vacuum initial data in ref. \cite%
{Mantoulidis:2014sba}. Following suggestions in \cite%
{O'Murchadha:2009kc,Malec:2015oza}, one may then reformulate the hoop
conjecture as 
\begin{equation}
\beta _{b}\leq 2\pi GM_{BY}/c^{2},  \label{hoop2}
\end{equation}%
where $M_{BY}$ is the Brown-York quasi-local mass \cite%
{Brown:1991gb,Brown:1992br} of the apparent horizon. Note that the the
Brown-York quasi-local mass is only defined for time-symmetric data. Using a
result of Paiva \cite{Paiva} \footnote{%
There is an earlier and weaker result due to Croke \cite{Croke2} which may
possibly prove to be of use in the present context}. One may check that (\ref%
{hoop2}) holds for all initial data sets constructed in ref. \cite%
{Mantoulidis:2014sba}. For a proof in Robinson-Trautmann metrics, see ref 
\cite{sken}.

The Brown-York quasi-local mass \cite{Brown:1991gb,Brown:1992br} of the
apparent horizon, which is assumed to have positive Gaussian curvature,
therefore admits a unique (up to rigid motions) isometric embedding into
Euclidean space as a convex body. The definition of the Brown-York mass
inside any 2-surface, $S,$ lying in a Cauchy surface $\{\Sigma ,\hat{g}\}$
is 
\begin{equation}
M_{BY}=\frac{1}{8\pi }\int_{S}\Bigl (k_{0}-k\Bigr )dA(S,\hat{g}|_{S}),
\end{equation}%
where $k$ is the trace of the fundamental form of $S$ considered as embedded
in $\{\Sigma ,\hat{g}\}$ and $k_{0}$ is the trace of the fundamental form of 
$\{S,g\}$ when isometrically embedded in Euclidean space $\{\mathbb{E}%
^{3},\delta _{ij}\}$, for which we have simply $dA(S,\hat{g}|_{S})=dA(S,{%
\delta _{ij}}_{S})=dA$. From a spacetime point of view, the Brown-York mass
depends on both how the spacelike surface $S$ sits in spacetime $\{M,g_{\mu
\nu }\}$ (it has two fundamental forms) and also the spacelike hypersurface $%
\Sigma $ passing through it (which picks out a linear combination of its two
second fundamental forms). The Brown-York mass is believed to be a
\textquotedblleft quasi-local\textquotedblright\ measure of the amount of
\textquotedblleft energy\textquotedblright\ on $\Sigma $ inside $S$ \cite%
{Szabados:2009eka}. The York-Brown mass suffers from a number of
shortcomings but in the present context has been shown that it is positive 
\cite{ShiTam0}.

Among the shortcomings of the Brown-York mass is that it requires that the
surface $S$ admit an isometric embedding into three-dimensional Euclidean
space. This is not possible for the horizon of all Kerr black holes.
Embeddings into four-dimensional Euclidean space are known but are believed
not to be unique. There exists a unique isometric embedding into hyperbolic
three-space \cite{Gibbons:2009qe} and hence a (presumably not unique)
embedding into four-dimensional Minkowski spacetime.

The converse of the hoop conjecture remains to be considered; that is, the
question if some surface $S$ satisfies

\begin{equation}
\beta _{b}\leq \ \frac{2\pi GM_{BY}}{c^{2}}\,,
\end{equation}%
then must $S$ be, or lie inside, an apparent horizon? The various papers of
Shi and Tam \cite%
{ShiTam0,ShiTam1,ShiTam2,ShiTam22,ShiTam3,ShiTam4,ShiTam5,ShiTam6} contain
some relevant results here.

\subsection{Relation to work of Tod}

Tod \cite{Tod} has looked at the hoop conjecture from the point of view of a
collapsing shell construction for which an isometric embedding is possible,
and seeks to define the hoop radius in terms of a \textit{maximum shadow
circumference }$C_{m}$. This is defined as the supremum of the circumference
of all orthogonal projections of the surface. He finds that 
\begin{equation}
\frac{\pi }{2}C\leq \frac{1}{2}\int_{\S }k_{0}dA\leq 2C_{m}\,,
\end{equation}%
where the upper bound is attained for any surface of constant breadth.

Thus, in the context of the time-symmetric initial value problem, rather
than the collapsing shell calculation \footnote{%
for which if the apparent horizon lies in a spacelike hyperplane, one does
have \cite{Gibbons:2009xm} 
\begin{equation}
\frac{\beta _{b}}{4\pi }\leq \frac{GM_{ADM}}{c^{2}}
\end{equation}%
.}, an apparent horizon must satisfy 
\begin{equation}
\frac{C_{m}}{8}\leq \frac{GM_{BY}}{c^{2}}\leq \frac{C_{m}}{2\pi }\,
\end{equation}%
Since 
\begin{equation}
\beta _{b}\leq C_{m}\,,
\end{equation}%
the lower bound yields 
\begin{equation}
\frac{\beta _{b}}{8}\leq \frac{GM_{BY}}{c^{2}}
\end{equation}%
which, since $8>2\pi $, is a weaker statement than (\ref{hoop2})\thinspace .

\section{\emph{\protect\bigskip }Conclusions}

We have explored the nature of a number of upper bounds on fundamental
quantities in nature. Some of this involves further elaboration and
generalisation to higher dimensions of earlier upper bounds on forces and
power in general relativity, but our discussion has focussed on a detailed
analysis of our conjecture that the ratio of the magnetic moment to angular
momentum is bounded above in nature. Suspicion falls on this combination for
a maximum principle because it has a natural Stoney-Planck unit that is
independent of the quantum of action, $h$, and so it entirely classical. We
find evidence for our conjecture that the ratio $c\mu /JG^{\frac{1}{2}}$ is
bounded by a quantity of order unity by investigating a wide range of
testing theoretical situations. In particular, we verified that such a
conjecture holds for charged rotating black holes in those theories for
which exact solutions are available, including the Einstein-Maxwell and
dilaton theories, Kaluza-Klein theory, the Kerr-Sen black hole, and the
so-called STU family of charged rotating supergravity black holes. We also
discussed the current status of the Maximum Tension Conjecture, the Dyson
Luminosity Bound, and Thorne's Hoop Conjecture and saw the possible points
of contact between them and our conjecture bounding $\mu /J$.

\textbf{Acknowledgements}

The authors are supported by the STFC of the United Kingdom. We would like
to thank Mirjam Cvetic, Chris Pope, Bruce Allen, Jorge Rueda, Michael Good
and Kostas Skenderis for helpful conversations and communications.

\end{document}